\documentstyle[prd,aps,eqsecnum,epsfig]{revtex}

\def\mev{\,{\rm MeV}}
\def\gev{\,{\rm GeV}}
\renewcommand{\c}{{\rm{c}}}

\renewcommand{\d}{\rm{d}}
\renewcommand{\u}{\rm{u}}
\newcommand{\q}{\rm{q}}
\newcommand{\s}{{\rm{s}}}

\newcommand{\qbar}{\overline{\rm{q}}}

\newcommand{\sbar}{\overline{\rm{s}}}
\def\cbc{\rm{c}\overline{\rm c}}
\def\qbq{{\rm q}\overline{\rm q}}
\def\sbs{{\rm s}\overline{\rm s}}
\def\ubu{{\rm u}\overline{\rm u}}
\def\dbd{{\rm d}\overline{\rm d}}
\def\jp{J/\psi}
\def\mjp{M_{J/\psi}}

\def\als{\alpha_s}
\def\sudrei{{\rm SU}(3)_{{\rm F}}}
\def\uaeins{{\rm U}(1)_{\rm A}}

\begin{document}

\draft
\preprint{
hep-ph/9802409,
Wuppertal WU B 98-2,
Heidelberg HD-THEP-98-5
}

\title{
Mixing and Decay Constants of Pseudoscalar Mesons\thanks{hep-ph/9802409,
Wuppertal WU B 98-2,
Heidelberg HD-THEP-98-5 (to be
published in {\it Physical Review}\/ D)}}

\author{Th.\ Feldmann\thanks{Supported by {\it Deutsche
      Forschungsgemeinschaft}},
    P.\ Kroll
}
\address{Fachbereich Physik, Universit\"at Wuppertal,
  D-42097 Wuppertal, Germany}
\author{ B.\ Stech
}
\address{ Institut f\"ur Theoretische Physik, Universit\"at Heidelberg,
  D-69120 Heidelberg, Germany}

\date{
{\tt Copyright 1998 by The American Physical Society.}}
\maketitle

\begin{abstract}
We propose a new $\eta$--$\eta'$ mixing scheme where we start from
the quark flavor basis and assume that the decay constants in that
basis follow the pattern of particle state mixing. On exploiting
the divergences of the axial vector currents --
which embody the axial vector anomaly --
all basic parameters are fixed to first order of flavor symmetry
breaking. That approach naturally leads to a mass matrix, 
quadratic in the masses, with specified elements. We also test
our mixing scheme against experiment and determine corrections
to the first order values of the basic parameters from 
phenomenology. Finally,
we generalize the mixing scheme to include the $\eta_c$.
Again the divergences of the axial vector currents fix the
mass matrix and, hence, mixing angles and the charm content
of the $\eta$ and $\eta'$.
\end{abstract}



\section{Introduction}

$\eta$--$\eta'$ mixing is a subject of considerable interest that has 
been examined in many investigations, see
e.g.\ \cite{Fritzsch,Isgur,Gilman:1987ax,dmitrasinovic97} 
and references therein. As 
it is well-known \cite{Witten}, the $\uaeins$ anomaly plays a decisive
role. For the octet-singlet mixing angle $\theta$ of these
pseudoscalar mesons values in the range of $-10^{\circ}$ to
$-23^{\circ}$ have been obtained depending on details of the analysis,
see  e.g.\ \cite{Gilman:1987ax}. The phenomenological analysis often 
involves decay processes where, besides state mixing, also weak decay
constants appear. The decay constants are defined by
\begin{equation}
\langle 0 | J_{\mu5}^i | P (p)\rangle = \imath \,
f_P^i \, p_\mu  \qquad (i=8,1; \quad P=\eta, \eta')\, ,
\label{dec}
\end{equation}
where $ J_{\mu5}^8$ denotes the $\sudrei$ octet and $ J_{\mu5}^1$ the 
$\sudrei$ singlet axial-vector current, respectively. 
Frequently, it is assumed that the decay constants follow the
pattern of state mixing: 
\begin{eqnarray}
&& f_{\eta\phantom{'}}^8 = f_8 \, \cos\theta \ , \qquad
   f_{\eta\phantom{'}}^1 = - f_1 \, \sin\theta \ , \cr
&&  f_{\eta'}^8 = f_8 \, \sin\theta \ , \qquad
   f_{\eta'}^1 = \phantom{-} f_1 \, \cos\theta \ .
\label{oldmix}
\end{eqnarray}
However, recent theoretical \cite{Leutwyler97} as well as  
phenomenological \cite{FeKr97b} investigations have shown that 
Eq.~(\ref{oldmix}) cannot be correct.
Adopting the new and general parametrization \cite{Leutwyler97}
\begin{eqnarray}
&& f_{\eta\phantom{'}}^8 =  f_8 \, \cos\theta_8 \ , \qquad
    f_{\eta\phantom{'}}^1 = - f_1 \, \sin\theta_1 \ ,\cr
&&  f_{\eta'}^8 = f_8 \, \sin\theta_8 \ , \qquad
   f_{\eta'}^1 = \phantom{-} f_1 \, \cos\theta_1 \ .
\label{newmix}
\end{eqnarray}
$\theta_1$ and $\theta_8$ turned out to differ considerably.
The phenomenological analysis  \cite{FeKr97b},
which involved the combined analysis of
the two-photon decay widths of the $\eta$ and $\eta'$,
the $\eta\gamma$ and $\eta'\gamma$ transition form factors
and the additional constraint from the radiative $\jp$ decays, 
allowed to determine the four quantities occuring in Eq.~(\ref{newmix}).
Most importantly, the values obtained satisfy the constraints 
\cite{Leutwyler97} from chiral perturbation theory (ChPT).

The appearance of the four parameters $f_8,f_1,\theta_8,\theta_1$
raises anew the problem of their mutual relations and their 
connection with the mixing angle of the particle states. This angle
is necessarily a single one since mixing with higher states --
the $\eta_{\c}$ for instance -- can be neglected at this stage.
The relation of this angle with the four parameters 
in Eq.~(\ref{newmix}) does not need to be simple: in a parton language 
\cite{BL80} the decay constants are controlled by specific 
Fock state wave functions at zero spatial separation of the 
quarks while state mixing refers to the mixing in the overall 
wave functions.

In this work we express $\eta$ and $\eta'$ as linear combinations of 
orthogonal states $\eta_{\q}$ and $\eta_{\s}$ which can be generated 
by the axial vector currents with the flavor structure 
$\qbq=(\ubu + \dbd)/\sqrt2$ and $\sbs$, respectively. They
are chosen in such a way that both states have vanishing
vacuum-particle matrix elements with the opposite currents, i.e.\
their lowest Fock space components have the compositions $\qbq$ and $\sbs$,
respectively. 
Our motivation for choosing this specific basis comes 
from the fact that the breaking of $\sudrei$ by the quark masses 
influences the two parts differently, and from the observation that
vector and tensor mesons -- where the
axial vector anomaly plays no role -- have state mixing angles
very close to the ideal mixing angle $\theta_{ideal}= \arctan \sqrt{2} $.
We will demonstrate that the proper use of this  quark flavor 
basis provides for new insights and successful
predictions.
We point out that we employ fixed (momentum-independent)
basis states. Thus, our state mixing angle is
momentum independent and well-defined also in any other basis
obtained by an orthogonal transformation. This differs from
other possible approaches in which momentum dependent mass
matrices are introduced (see, for instance, \cite{JuWe98}).
The decay constants, on the other hand, will in general depend on $q^2$,
i.e.\ the particle states and masses, and will thus require a
parametrization by two different mixing angles as in (\ref{newmix}).
These angles depend on the basis which is used for the definition of the
decay constants. As described below, the basic assumption which we will use
in this paper is that the decay constants follow the state mixing
if and only if they are defined with respect to the quark basis.
In this circumstance the two angles for the
decay constants obtained in this basis and the corresponding state
mixing angle coincide and are thus momentum-independent.

Defining now decay constants analogous to Eq.~(\ref{dec}) but with 
$ i=\q, \s$ and denoting the 
$\eta$~--~$\eta'$ mixing angle that describes the deviation from 
ideal mixing, by $\phi$, we propose
\begin{eqnarray}
&& f_\eta^{\q} = f_{\q} \, \cos\phi \ , \qquad
    f_\eta^{\s} = - f_{\s} \, \sin\phi \ ,\nopagebreak \cr\nopagebreak 
&&  f_{\eta'}^{\q} = f_{\q} \, \sin\phi \ , \qquad
   f_{\eta'}^{\s} =  \ \ f_{\s} \, \cos\phi \ .
\label{phimix}
\end{eqnarray}
That the decay constants in the quark flavor basis follow in this way
the pattern of particle state mixing is our central assumption. It is
equivalent to the requirement that the contribution $f_{\q}$ ($f_{\s}$) 
to the decay constants obtained
from the $\eta_{\q}$ ($\eta_{\s}$) components of the wave functions
is independent of the meson involved. This assumption appears plausible
but we have no rigorous justification for it and have to test it. It is
certainly restrictive as will be shown
in Sect.~\ref{theorysect}: first, it reduces the number of parameters again
to three. Secondly, by invoking the divergences of the currents, the
angle $\phi$ is connected to $f_{\q}/f_{\s}$.
Finally, flavor symmetry fixes $f_{\q}$ and $f_{\s}$
to first order of $\sudrei$ breaking, leaving us -- to this order -- with
no free parameter. Mass mixing of the pseudoscalar mesons is also
discussed in this section.
Numerous phenomenological checks are
possible and performed in Sect.\ \ref{phensect}. We determine 
phenomenological values for the
three parameters from the data and check for consistency
with ChPT and the earlier determination 
\cite{FeKr97b} of the four quantities  $f_8,f_1,\theta_8,\theta_1$.
Our scheme will then be generalized to include the $\eta_{\c}$ in Sect.\
\ref{etacsect}. 
The generalized approach allows to estimate the quark content of
the three pseudoscalar mesons, the mixing angles and the charm decay
constants $f^{\c}_{\eta}$ and $f^{\c}_{\eta'}$ which attracted much
interest in the current discussion \cite{Cheng,Halperin,AlGr97,Ali97b}
of the rather large branching ratio for the process $B\to K\eta'$ as 
measured by CLEO \cite{exp}. Our summary is presented in Sect.~\ref{summsec}.


\section{The $\qbq$~--~$\sbs$ mixing scheme}
\label{theorysect}

The two states $\eta_{\q}$ and $\eta_{\s}$
are related
to the physical states by the transformation
\begin{equation}
 \left (\matrix{\eta \cr \eta'}\right )\,=\, U(\phi)\;
                      \left (\matrix{\eta_{\q} \cr \eta_{\s}}\right ) ,
\label{qsb}
\end{equation}
where $U$ is a unitary matrix defined by
\begin{equation}
    U(\alpha)\,=\,\left(\matrix{\cos{\alpha} & -\sin{\alpha} \cr
                                \sin{\alpha} & 
\phantom{-}\cos{\alpha}} \right) .
\label{uni}
\end{equation}
We assume that the physical states are orthogonal, i.e.\ that  
mixing with
heavier pseudoscalar mesons (e.g.\ the
$\eta_{\c}$) can be ignored, see, however, Sect.~\ref{etacsect}. 
We stress that as long
as state-mixing is considered, one may freely transform from
one orthogonal
basis to the other. 
For example, the standard octet-singlet mixing angle is
given by $\theta=\phi - \theta_{\rm ideal}$.
According to our 
central assumption described in Sect.\ 1, we take
\begin{equation}
\left(\matrix{f_{\eta}^{\q} & f_{\eta}^{\s} \cr f_{\eta'}^{\q} &
                        f_{\eta'}^{\s}}\right) \,=\,
                                U (\phi)\; {\mathcal F} \ ,
\quad {\mathcal F}\,=\, \left(\matrix{f_{\q}& 0 \cr 0 & f_{\s}}\right)  .
\label{phqs}
\end{equation}
Transforming the non-strange and strange axial-vector currents into octet and
singlet currents, one can also connect the decay constants defined in
Eq.~(\ref{dec})
to $f_{\q}$ and $f_{\s}$
\begin{equation}
\left(\matrix{f_{\eta}^8 & f_{\eta}^1 \cr f_{\eta'}^8 &
                        f_{\eta'}^1}\right) \,=\,
          U(\phi)\;{\mathcal F}\;U^{\dagger}(\theta_{ideal})
\label{phos}
\end{equation}
with the result
\begin{eqnarray}
&& f_8 = \sqrt{1/3 \, f_{\q}^2 + 2/3 \, f_{\s}^2} \ , \qquad
   \theta_8 = \phi - \arctan (\sqrt{2}\,f_{\s}/f_{\q})  \ , \cr
&& f_1 = \sqrt{2/3 \,f_{\q}^2 + 1/3 \, f_{\s}^2}  \ , \qquad
   \theta_1 = \phi - \arctan (\sqrt{2}\,f_{\q}/f_{\s}) \ 
\label{fleu}
\end{eqnarray}
and thus 
\begin{eqnarray}
         \tan (\theta_1-\theta_8) = \sqrt{2}/3 \, 
(f_{\s}/f_{\q} - f_{\q}/f_{\s}) ~ .
\label{theta}
\end{eqnarray}
These results clearly show that, as a 
consequence of $\sudrei$ breaking, 
at most for a single choice of the basis states
the matrix of the decay constants follows
the particle state mixing as in
Eq.~(\ref{phqs}). For the reasons mentioned in the introduction
we assume Eq.~(\ref{phqs}) to hold in the $\qbq$-$\sbs$ basis.
It implies that the decay constants of the mesons are mass
independent superpositions of $f_{\q}$ and $f_{\s}$.
The difference between our ansatz for the decay constants
and the customary ones lies in the treatment of $\sudrei$
breaking effects, which naturally manifest themselves in
the ratio $f_{\q}/f_{\s} \neq 1$.
In order to proceed we consider the divergences of the axial vector
currents. They embody the well-known axial vector anomaly, for instance
\begin{eqnarray}
\partial^\mu J_{\mu 5}^{\s} =
\partial^\mu (\bar{\s} \, \gamma_\mu \gamma_5 \, \s) &=&
2 \, m_{\s} \, (\bar{\s} \, i \gamma_5 \, {\s}) + \frac{\alpha_s}{4\pi} \,
G \, \tilde G \ .
\end{eqnarray}
$G$ denotes the gluon field strength tensor and $\tilde G$
its dual; $m_i$ denote the current quark masses.
The vacuum--meson transition matrix elements of the axial
vector current divergences are given by the product of the
square of the meson mass, $M_P^2$, and the appropriate decay
constant. For instance,
\begin{equation}
\langle 0 | \partial^\mu J_{\mu 5}^{\s} | \eta \rangle = M_\eta^2 \,
f_\eta^{\s}
\ .
\end{equation}
The mass factors, which necessarily appear quadratically here,
can be considered as the elements of the particle mass matrix
\begin{equation}
{\mathcal M}^2 = \left(\matrix{ M_\eta^2 & 0 \cr 0 & M_{\eta'}^2 } \right)
\ .
\end{equation}
With the help of Eq.~(\ref{phqs}) the matrix elements of $\partial^\mu
J_{\mu 5}^i$ ($i=q,s$) can then be identified as those of the matrix
product ${\mathcal M}^2 \, U(\phi) \, {\mathcal F}$. Transforming it to
the quark flavor basis and solving for the mass matrix
\begin{equation}
{\mathcal M}^2_{\q\s} = U^\dagger(\phi) \, {\mathcal M}^2 \, U(\phi)
\label{massrot} \ ,
\end{equation}
one easily finds
\begin{equation}
{\mathcal M}^2_{\q\s} = \left(\matrix{
 m_{\q\q}^2 
 + \frac{\sqrt2}{f_{\q}}\,
\langle 0|\frac{\alpha_s}{4\pi}\,G\tilde G|\eta_{\q}
  \rangle &
 \frac{1}{f_{\s}}\,\langle 0|\frac{\alpha_s}{4\pi}\,G\tilde G|\eta_{\q}
  \rangle \vspace{0.3em}\cr
 \frac{\sqrt2}{f_{\q}}\,\langle 0|\frac{\alpha_s}{4\pi}\,G\tilde G|\eta_{\s}
  \rangle &
 m_{\s\s}^2 
 + \frac{1}{f_{\s}}\,\langle 0|\frac{\alpha_s}{4\pi}\,G\tilde G|\eta_{\s}
  \rangle } \right)
\label{qsmass}
\end{equation}
where we use the abbreviations
\begin{eqnarray}
&& m_{\q\q}^2 = \frac{\sqrt2}{f_{\q}} \, 
      \langle 0 | m_{\u} \, \bar{\u} \, i \gamma_5 \, {\u} +
                  m_{\d} \, \bar{\d} \, i \gamma_5 \, {\d} | \eta_{\q}
                  \rangle;
\quad
m_{\s\s}^2 = \frac{2}{f_{\s}} \, 
      \langle 0 | m_{\s} \, \bar{\s} \, i \gamma_5 \, {\s} | \eta_{\s}
                  \rangle
\label{miidef}
\end{eqnarray}
for the quark mass contributions to ${\mathcal M}_{\q\s}^2$.
As expected, the anomaly is the only source of the non-diagonal
elements. The symmetry of the mass matrix forces an important 
connection between the ratio of couplings of our basis states to
the anomaly
\begin{equation}
y = \sqrt2 \, \frac{ \langle 0 | 
                  \frac{\alpha_s}{4\pi} \, G \tilde G | \eta_{\s} \rangle}
{\langle 0 | 
                  \frac{\alpha_s}{4\pi} \, G \tilde G | \eta_{\q} \rangle}
= \frac{f_{\q}}{f_{\s}} \ . 
\label{ydef}
\end{equation}
For later use it is also convenient to introduce the
abbreviation
\begin{equation}
a^2 = \frac{1}{\sqrt2 \, f_{\q}} \, \langle 0 | 
                  \frac{\alpha_s}{4\pi} \, G \tilde G | \eta_{\q}
                  \rangle
\label{aadef}
\end{equation}
for the anomaly contribution to the mass matrix. On account of
Eqs.~(\ref{massrot}) and (\ref{qsmass}) both, $a^2$ and $y$, can be 
expressed in terms of the masses and the mixing angle $\phi$
\begin{eqnarray}
a^2 &=& \frac{M_\eta^2 \, \cos^2\phi + M_{\eta'}^2 \, \sin^2\phi -
  m_{\q\q}^2}{2}\, ,
\label{aavalue}
\\[0.3em]
y &=&  \frac{(M_{\eta'}^2-M_{\eta}^2)\, \sin 2\phi}
{2\sqrt2 \, a^2}\, .
\label{yvalue}
\end{eqnarray}
The combination of Eqs.~(\ref{ydef}) and (\ref{yvalue}) 
provides an interesting relation between $f_{\q}/f_{\s}$
and the mixing angle $\phi$. 
Also worth noting is the relation between $\phi$ 
and $\theta_8$ that is obtained  by
combining Eq.~(\ref{fleu}) with Eqs.~(\ref{ydef}) and (\ref{yvalue})
\begin{equation}
\cot \theta_8\,=\,-\, \frac{M_{\eta'}^2}{M_{\eta}^2} \, \tan\phi .
\label{yphi}
\end{equation}
This relation holds up to corrections of order $m_{\q\q}^2/M^2_P$.

Flavor symmetry allows to relate $m_{\q\q}^2$ and $m_{\s\s}^2$, 
defined in Eq.~(\ref{miidef}), to the pion and kaon masses. 
To the order we are working, one gets
\begin{equation}
m_{\q\q}^2 = M_\pi^2 \ , \quad
m_{\s\s}^2 = 2 \, M_K^2 - M_\pi^2 \ .
\label{masses}
\end{equation}
These relations allow for a first and -- to the given order -- 
parameter-free
application of our scheme for the determination of all quantities
relevant for $\eta$--$\eta'$ mixing provided the values of the physical
particle masses are given. Using Eqs.~(\ref{aavalue}) and (\ref{yvalue})
 as well as
Eq.~(\ref{masses}) for the corresponding elements of the mass matrix 
(\ref{qsmass}), we evaluate 
$\phi$, $a^2$ and $y$, and hence $\theta$. 
The results for these quantities are listed in Table~\ref{comptable1}.

For a theoretical estimate of $f_{\q}$ and $f_{\s}$ we take over
their particle independence which was necessary for Eq.~(\ref{phimix}) to hold, 
to the $\pi$ and the $K$ meson. We retain the difference between the 
pion and kaon decay constants as a first order correction due to
flavor symmetry breaking: 
 \begin{equation}
f_{\q} = f_\pi \ , \qquad
f_{\s} = 
    \sqrt{2 \, f_K^2 - f_\pi^2}  ~ .
\label{phen6}
\end{equation}
Note that $V$-spin considerations provide a linear relation
between the decay constants $f_{\s}$, $f_\pi$ and
$f_K$ which, to the considered order of flavor symmetry breaking,
can be replaced by the above quadratic relation.
As can be seen from Eq.~(\ref{theta}), these theoretical results for
$f_{\q}$ and $f_{\s}$ lead to 
a substantial difference between $\theta_1$ and
$\theta_8$. Only in the strict $\sudrei$ limit where 
$f_{\q}=f_{\s}$ (or, equivalently, $f_K=f_\pi$ or $f_8=f_1$) 
one would have $\theta_1=\theta_8=\theta$, with $\theta$
being the octet-singlet mixing angle.
According to Leutwyler \cite{Leutwyler97}, ChPT provides two relations
(up to $1/N_c$ corrections) among the decay constants
\begin{eqnarray}
&& f_8 = \sqrt{\frac43 \, f_K^2 - \frac13 \, f_\pi^2} 
\label{f8rel} \ , \qquad
f_{\eta}^8 \, f_{\eta}^1 + f_{\eta'}^8 \, f_{\eta'}^1 =
- \frac{2\sqrt2}{3} \, (f_K^2-f_\pi^2).
\label{anglediffex}
\end{eqnarray}
It can easily be verified that, on using Eq.~(\ref{phen6}), these
relations are satisfied in our approach.
By means of Eq.~(\ref{fleu}) the theoretical values of $f_8$, $f_1$, $\theta_8$ 
and $\theta_1$ are determined by the decay constants presented in 
Eq.~(\ref{phen6}) and the mixing angle $\phi$ computed from the mass
matrix (see Table \ref{comptable2} below). 
The numerical values of the mixing parameters, resulting from 
Eqs.~(\ref{masses}),(\ref{phen6}), may, of course, be subject to sizeable
corrections (of $O(1/N_c)$ in the language of ChPT).
As an example of the size of such corrections we note that from
Eqs.~(\ref{phen6}) and (\ref{ydef}) $y=0.71$ follows which differs from the
value obtained from the mass matrix, see Table \ref{comptable1}. 
The corrections to the mixing parameters will phenomenologically be
estimated in the next section.  

The considerations presented in this section nicely
demonstrate that our approach is indeed very restrictive. To the
order of flavor symmetry breaking we are working, there is no
free parameter left. As is to be emphasized this interesting
outcome crucially depends on the central assumption (\ref{phimix}).
If, in analogy to Eq.~(\ref{newmix}), we allowed for two
angles in the quark flavor basis for the parametrization of
the decay constants, our approach would loose its predictive power
completely\footnote{
Using the phenomenological parameters found in \cite{FeKr97b},
and translating them into the quark flavor basis with admission of two
mixing angles, $\phi_{\q}$ and $\phi_{\s}$, defined in analogy to
Eq.~(\ref{newmix}), one finds
$f_{\q} = 1.09 \, f_\pi$; $f_{\s}= 1.38 \, f_\pi$; $\phi_{\q}=
39.4^\circ$ and $\phi_{\s}= 38.5^\circ$. The fact that
the two angles nearly coincide give direct support to the validity
of Eq.~(\ref{phimix}).}. 
In the next section we will confront our approach with experiment. 
We determine the mixing angle $\phi$ and the basic decay constants 
$f_{\q}$ and $f_{\s}$ phenomenologically
and look for consistency and for deviations from the first order 
of $\sudrei$ breaking.


\section{The phenomenological values of $\phi$, $f_{\q}$, $f_{\s}$}
\label{phensect}

\begin{figure}[thb]
\begin{center}
\epsfclipon
\psfig{file = 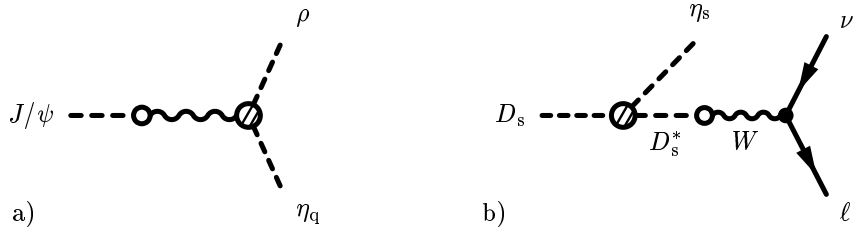, bb =  110 645 440 745}
\end{center}
\caption{a) Electromagnetic contribution to $J/\psi \to \rho
  \eta,\eta'$. b) Pole ansatz for the $D_{\s}\to P \ell \nu$
semi-leptonic decay.}
\label{fig}
\end{figure}

Several possibilities to extract the value of
the mixing angle from experiment have been discussed in the
literature, see, for instance, 
\cite{Fritzsch,Gilman:1987ax,BaFrTy95,Bramon97}.
We can profit from these papers by properly adapting them to the 
$\q\qbar-\s\sbar$ mixing scheme. 
We note that in the phenomenological analyses
\cite{Fritzsch,Gilman:1987ax,BaFrTy95,Bramon97} some additional
simplifying assumptions had to be made. 
Thus, for instance,
OZI-suppressed contributions or mass dependencies of 
form factors and coupling constants are ignored.
We start by first discussing processes which are independent 
(or insensitive) to the decay constants and allow a one parameter 
fit of the particle state mixing angle $\phi$.
\begin{itemize}
\item  {\it The decay $J/\psi \to P\rho$}: We consider the ratio 
of the decay
widths $\Gamma[J/\psi \to \eta'\rho]$ and $\Gamma[J/\psi \to
\eta\phantom{{}'}\rho]$. 
In these processes $G$-parity is not conserved,
they proceed through a virtual photon (see Fig.\ \ref{fig}a). 
Contributions
{}from the isospin-violating part of QCD are supposedly very 
small as
can be inferred from the smallness of the $\jp\to\phi\pi$ width 
and will be
neglected. The calculation of the decay widths  
requires the knowledge of the $\rho$--$P$ transition form factors at 
momentum transfer $q^2=M_{J/\psi}^2$. 
On account of the flavor content of the $\rho$ meson, this transition
form factor only probes the $\eta_{\q}$ components of the $\eta$ and
$\eta'$ if OZI-suppressed contributions are neglected. Hence,
\begin{eqnarray}
F_{\rho\eta\phantom{{}'}}(q^2) &=& \cos\phi \, F_{\rho\eta_{\q}}(q^2), \cr
F_{\rho\eta'}(q^2) &=& \sin\phi \, F_{\rho\eta_{\q}}(q^2)\, .
\label{PrhoFF}
\end{eqnarray}
and therefore 
\begin{eqnarray}
\frac{\Gamma[J/\psi \to \eta'\rho]}
     {\Gamma[J/\psi \to \eta\phantom{{}'}\rho]}  = \tan^2 \phi \, 
\left(\frac{
    k_{\eta'\rho}}{k_{\eta\phantom{{}}\rho}}\right)^3
\label{rpv}
\end{eqnarray}
where 
\begin{equation}
k_{PV}= \mjp \,[1-(M_P^2+M_V^2)/\mjp^2]\,/2.
\label{mom}
\end{equation}
From the experimental
value  $0.54 \pm 0.11$ for this ratio of decay widths  \cite{PDG96}
we obtain $\phi = 39.9^\circ \pm 2.9^\circ$. 
Almost the same value for $\phi$ has been found in an analysis 
of all isospin-1 $\jp\to PV$ decays (including pions)\cite{Bramon97}.
A global fit to all $\jp\to PV$ 
decay modes on the basis of a particular model, yields 
$\phi = 37.8^{\circ}\pm 1.7^{\circ}$\cite{Bramon97}. 
Because of its model dependence we will not use the latter
result in evaluating the average of the mixing angle.
Since in the derivation of Eq.~(\ref{rpv}) only the mixing angle of the 
particle states enters, one may freely transform from the 
$\q\qbar-s\sbar$ basis to the octet-singlet one as was done, 
for instance, in \cite{Bramon97}. Nevertheless, 
the simple relation between the ratio of decay widths and the mixing angle,
independent of the dynamics, is an advantage of the $\q\qbar-\s\sbar$ basis 
used here. In the octet-singlet mixing scheme one would have to deal 
with a linear combination of two a priori different form factors. We will
profit from this advantage also in the following five processes.
\item {\it The decays $\eta' \to \rho \gamma$ and $\rho \to \eta \gamma$}:
The transition matrix elements controlling these processes can be
decomposed covariantly \cite{Dumbrajs:1983jd}
\begin{equation}
\langle \gamma P(p_P)|T| \rho (p_{\rho})\rangle 
            = -e g_{\rho P\gamma} \, \epsilon_{\mu\nu\lambda\sigma}
                 \, p_P^\mu \, \varepsilon_\gamma^{(*)\nu}\, 
                  p_{\rho}^\lambda \varepsilon_{\rho}^{\sigma}\, ,
\label{gvp}
\end{equation}
leading to the following expressions for the decay widths
\begin{eqnarray}
&& \Gamma[ \eta' \to \rho \gamma ] = \alpha \, g_{\rho \eta'\gamma}^2 \, 
      k_{\rho}^3 \ , \qquad
\Gamma[ \rho \to \eta \gamma ] = \frac{\alpha}{3} \, 
g_{\rho \eta\gamma}^2 \, 
              k_{\eta}^3 ~.
\label{PVgamma} 
\end{eqnarray}
where
\begin{equation}
k_{\rm f}=M_{\rm i} \,[1- M_{\rm f}^2/M_{\rm i}^2]/2 \ , 
\label{momentum}
\end{equation}
being the 3-momenta of the final state meson~f.
$M_{\rm i}$ denotes the mass of the decaying meson. 
$\alpha$ is the fine structure constant.
Using state mixing
(\ref{qsb}), one finds 
\begin{equation}
\frac{g_{\rho\eta'\gamma}}{g_{\rho \eta\gamma}}\,=\,\tan{\phi}\, .
\label{pv}
\end{equation}
From the measured decay widths \cite{PDG96} we obtain for the ratio of
the $\rho \eta (\eta')\gamma$ coupling constants the value 
$1.41\pm 0.29$ from which $\phi =  35.3^{\circ}\pm 5.5^{\circ}$ 
follows\footnote{One may extend
this analysis to the  $\omega$ and $\phi$ cases. Ignoring the small
effect due to the $\omega-\phi$ mixing, one derives 
$g_{\omega\eta'\gamma}/g_{\omega\eta\gamma} \simeq \tan{\phi}$ and 
$g_{\phi\eta'\gamma}/g_{\phi\eta\gamma} \simeq \cot{\phi}$,
respectively. From the measured values/bounds \cite{PDG96}
we obtain $\phi \simeq 37^\circ \pm 8^\circ$ and $\phi > 21^\circ$,
respectively.}. 
\item {\it The decays $T \to P_1 P_2$}: 
Here $T$ denotes a $2^{++}$ tensor meson
and $P_i$ refers to a pseudoscalar meson.
Most suitable for a rather model-independent determination of $\phi$
are the $\eta (\eta') \pi$ decay modes of the $a_2$. Using the same 
assumptions as before one gets
\begin{eqnarray}
 \frac{\Gamma[a_2 \to \eta'\pi]}{\Gamma[a_2 \to \eta\phantom{{}'}\pi]}
&=&
\tan^2\phi \,
\left(\frac{k_{\eta'\pi}}{k_{\eta\phantom{{}'}\pi}}\right)^5 
     \, ,
\end{eqnarray}
where $k_{P\pi}$ is defined analogously to Eq.~(\ref{mom}).
{}From the experimental value for that ratio, 
$0.039 \pm 0.008$ \cite{PDG96}, we obtain
$\phi = 43.1^\circ\pm 3.0^\circ$.
In \cite{Bramon97} the whole class of $T \to P_1P_2$ decays has been  
analyzed in a model-dependent way recently. An overall fit to the data
is consistent with $\phi\simeq 39^\circ$.
\item {\it The decay $D_{\s} \to Pe\nu$}: The ratio of decay widths 
$\Gamma[D_{\s} \to \eta'e\nu]/\Gamma[D_{\s} \to \eta e\nu]$
is determined by the $D_{\s} \to \eta'$, $\eta$ form factors   
$f_+^{\eta'}(q^2)$ and $f_+^{\eta}(q^2)$. Using a pole ansatz for 
their $q^2$ dependence \cite{BSW} one can extract from the 
decay rates the form factor ratio at $q^2$= 0 which -- 
in our scheme -- is simply equal to $\cot \phi$. The analysis 
\cite{BaFrTy95} using a monopole behavior with the $D_{\s}^*$ pole see 
Fig.\ \ref{fig}b and CLEO data \cite{CLEODs}   gives the value 
of $1.14 \pm 0.17 \pm 0.13$
and, hence, $\phi=41.3^\circ \pm 5.3^\circ$.
\item {\it The scattering processes $\pi^-p \to Pn$}:
At high energies the ratio of the cross-sections
should be independent of phase-space corrections and is
given by
\cite{Gilman:1987ax,Bramon97} 
\begin{eqnarray}
 \frac{\sigma(\pi^-p \to \eta' n)}{\sigma(\pi^-p \to \eta\phantom{{}'}
   n)}
&=& \tan^2 \phi \quad (s \gg M_P^2)
\end{eqnarray}
The two experiments lead to $\phi=36.5^\circ\pm 1.4^\circ$
\cite{ape} and $39.3^\circ\pm1.2^\circ$ \cite{sta}. Since the two
results are not fully consistent with each other we will double
the errors in the evaluation of the averaged value $\bar\phi$. 
\item {\it Annihilation processes $p\bar{p}\to P M$} ($M=\pi^0,\eta,\omega$): 
The Crystal Barrel Collaboration \cite{ams} measured 
the ratios for annihilation into $\eta M$ and $\eta'M$ and quoted
a value of $\phi=37.4^\circ\pm 1.8^\circ$ for the mixing angle.
However, since the experiment was carried through at low energies,
the result for $\phi$ is rather sensitive to phase space factors
and to the momentum dependence of the annihilation amplitudes.
We therefore discard that value of $\phi$ in the determination
of the averaged mixing angle although it will turn out to be
consistent with it.
\item {\it The decay $J/\psi \to P\gamma$}:
According to \cite{Novikov,BaFrTy95} 
the photon is emitted by the $\c$ quarks which then 
annihilate into lighter quark pairs through the effect of the
anomaly. 
Thus, the creation of the corresponding light 
mesons is controlled by the matrix element  $\langle
0|\frac{\als}{4\pi} G\tilde{G}|P\rangle$. 
The photon emission from light quarks 
is negligibly small as seen from the smallness of the $\pi \gamma $ 
decay branching ratio. Using Eqs.~(\ref{qsmass}, \ref{ydef}) and (\ref{yphi}) 
as well as setting $m_{u,d} =0$, we have 
\begin{eqnarray}
R_{J/\psi}=\frac{\Gamma[J/\psi \to \eta'\gamma]}{
                 \Gamma[J/\psi \to \eta\phantom{{}'}\gamma]}
&=& \tan^2\phi \,
\frac{M_{\eta'}^4}{M_{\eta\phantom{{}'}}^4} \,
\left(\frac{k_{\eta'}}{k_{\eta\phantom{{}'}}}\right)^3
= \cot^2\theta_8 \, \left(\frac{k_{\eta'}}{k_{\eta\phantom{{}'}}}\right)^3
\label{rjpsi} \ .
\end{eqnarray} 
From the measured value \cite{PDG96} $R_{J/\psi}=5.0 \pm 0.6$ 
the mixing angle $\phi$ becomes $39.0^\circ\pm 1.6^\circ$.
Obviously,
Eq.~(\ref{rjpsi}) is not equivalent to the naive singlet dominance 
prediction for which the factor $\cot\theta_8$ would have to be replaced by
$\cot\theta$. As we learned in Sect.~\ref{theorysect}, $\theta_8$
markedly differs from the octet-singlet mixing angle.
\end{itemize}

A weighted average of the above seven  high-lighted values
yields
\begin{eqnarray} 
\overline {\phi} = 39.3^\circ \pm 1.0^\circ.
\label{ave}
\end{eqnarray} 
Quite remarkably, the values for $\phi$ obtained from
very different physical processes are all compatible with
each other within the errors. This is not the case in the octet-singlet 
scheme
($\theta_8=\theta_1=\theta$)
where values varying from $-10^\circ$ to $-23^\circ$
have been found for $\theta$ 
\cite{Gilman:1987ax,BaFrTy95,Bramon97}. The
phenomenological value of $\phi$ does not differ substantially from the
leading order value, i.e.\ the higher order flavor symmetry
breaking corrections, 
absorbed in the phenomenological value, are apparently not large
(see Table~\ref{comptable1}).  

Having fixed the mixing angle, we are in the position to 
determine phenomenologically the ratio of the decay constants 
$f_{\q}$ and $f_{\s}$ by combining Eqs.~(\ref{ydef}) and (\ref{yvalue}). We find, 
with $m_{qq}^2 = M_\pi^2$,
\begin{equation}
f_{\q}/f_{\s} = y = 0.81 \pm 0.03.
\end{equation}
It is amusing to note that the replacement of the kaon mass by an
effective mass of $508$~MeV in the mass matrix introduced in 
Sect.~\ref{theorysect}
reproduces the phenomenological values of $\phi$, $a^2$ and $y$ exactly.

\begin{itemize}
\item {\it The decay $P \to \gamma\gamma$}: 
The two-photon decays of the $\eta$ and the $\eta'$ provide independent
information on the two decay constants. 
Expressing the PCAC results, see \cite{FeKr97b}
and references therein, for the two-photon decay widths of the
$\eta$ and $\eta'$ in terms of $\phi$, $f_{\q}$ and $f_{\s}$ 
($C_{\q}=5/9\sqrt2$, $C_{\s}=1/9$)
\begin{eqnarray}
\Gamma[\eta\phantom{'}\to\gamma\gamma] &=&
\frac{9\alpha^2}{16 \pi^3} \, M_{\eta\phantom{{}'}}^3 \,
\left[\frac{C_{\q} \, \cos\phi}{f_{\q}} -
      \frac{C_{\s} \, \sin\phi}{f_{\s}}\right]^2 \nonumber \ , \\[0.3em]
\Gamma[{\eta'}\to\gamma\gamma] &=&
\frac{9\alpha^2}{16 \pi^3} \, M_{{\eta'}}^3 \,
\left[\frac{C_{\q} \, \sin\phi}{f_{\q}} +
      \frac{C_{\s} \, \cos\phi}{f_{\s}}\right]^2 \ , 
\label{eq:gammapred}
\end{eqnarray}
and solving for $f_{\q}$ and $f_{\s}$, we arrive at
\begin{eqnarray}
f_{\q} &=& \frac{3 \, C_{\q}\,\alpha}{4 \, \pi^{3/2}} \,
 \left[\phantom{-} \cos\phi \, \sqrt{\Gamma[\eta\to\gamma\gamma]/M_\eta^3}
       +\sin\phi \, \sqrt{\Gamma[\eta'\to\gamma\gamma]/M_{\eta'}^3}
     \, \right]^{-1} , \nonumber \\[0.3em]
f_{\s} &=& \frac{3 \, C_{\s}\,\alpha}{4 \, \pi^{3/2}} \,
 \left[ -\sin\phi \, \sqrt{\Gamma[\eta\to\gamma\gamma]/M_\eta^3}
       +\cos\phi \, \sqrt{\Gamma[\eta'\to\gamma\gamma]/M_{\eta'}^3}
     \, \right]^{-1} .
\label{pho}
\end{eqnarray}
We evaluate Eq.~(\ref{pho}) with the mixing angle according to Eq.~(\ref{ave}) 
and the
following experimental values for the decay widths
$\Gamma[\eta\to\gamma\gamma]  = (0.51 \pm 0.026)~$keV and 
 $\Gamma[\eta'\to\gamma\gamma]  = (4.26 \pm 0.19)~$keV \cite{PDG96}.
The value of $0.324\pm 0.046$~keV, obtained from the Primakoff
production measurement of $\eta\to\gamma\gamma$, is
not included. It turns out that $f_{\s}$ is not well determined this
way, it acquires a rather large error $f_{\s}= (1.42 \, \pm 0.16) \, f_\pi $.
We therefore evaluate $f_{\s}$ also from $f_{\q}$ and the 
phenomenological value of the ratio $y$ and form the
weighted average of both values to find a more precise value for 
$f_{\s}$. By this means we obtain 
\begin{eqnarray}
f_{\q} = (1.07 \pm 0.02) \, f_\pi , \quad
f_{\s} = (1.34 \pm 0.06) \, f_\pi .
\label{fvalues}
\end{eqnarray}
These values for the basic decay constants differ 
from the theoretical values (\ref{phen6}) only mildly. Since, within the
errors, both the values of $f_{\s}$ determined here 
agree with each other, the
experimental values of the two-photon decay widths are well reproduced
by the parameter set (\ref{ave}), (\ref{fvalues}).
\end{itemize}

As an immediate test of the parameters (\ref{ave}) and (\ref{fvalues})
we compute the $P\gamma$ transition form factors along the lines 
described in detail in \cite{FeKr97b}.
We find excellent agreement between theory and experiment \cite{exp}.
The new results are practically indistinguishable from the fit
performed in \cite{FeKr97b} ($\chi^2/$d.o.f.\ is 28/34 as compared
to 26/33 in \cite{FeKr97b}). The form factor analysis is based on a
parton Fock state decomposition of the physical mesons. 
The wave functions of the valence Fock states,
providing the leading contribution to the form factor above
$Q^2 = 1$~GeV$^2$, are assumed to have the asymptotic form.
The values of these wave functions at the origin of configuration
space are related to the decay constants \cite{FeKr97b}.

A comparison between the theoretical and phenomenological values of
the mixing parameters is made in Table \ref{comptable1}. As can be
noticed there is no substantial deviation between both set of values,
i.e.\ higher order $1/N_c$ corrections, absorbed in the
phenomenological values, seem to be reasonably small.
In Table \ref{comptable2} we list the values of the parameters
defined in Eq.~(\ref{newmix}), i.e.\ in the parametrization
introduced by Leutwyler \cite{Leutwyler97}, as obtained from various sources.
The theoretical values of $f_8$, $f_1$, $\theta_8$ are computed
from the decay constants given in Eq.~(\ref{phen6}) and the theoretical
mixing angle listed in Table \ref{comptable1} while the
phenomenological values follow from Eqs.~(\ref{ave}) and (\ref{fvalues}). 
As can be seen the results obtained from the analyses performed in
this work and in \cite{Leutwyler97,FeKr97b}
agree rather well with each other.
The conventional analyses, e.g.\ \cite{Gilman:1987ax,BaFrTy95}, 
are not included in the table because the
difference between $\theta_8$ and $\theta_1$ is not 
considered.

\section{Generalizing to $\eta$--$\eta'$--$\eta_{\c}$ mixing}
\label{etacsect}
{}From the previous sections we learned that our
central assumption (\ref{phimix}) combined with the divergences
of the axial vector currents leads to a variety of
interesting predictions which compare well with experiment.
The reason for this success is likely the rather large 
difference between the current masses of the strange and the
up/down quarks. Since the charm quark mass is even heavier than the
strange one, it is tempting to generalize to the
$\qbq$--$\sbs$--$\cbc$ 
basis and to assume a similar behavior for the decay constants of the 
$\eta$--$\eta'$--$\eta_{\c}$ system in that basis. 
Then we can write
\begin{equation}
 \left( \begin{array}{ccc}
f_{\eta\phantom{{}'}}^{\q} & f_{\eta\phantom{{}'}}^{\s} 
& f_{\eta\phantom{{}'}}^{\c} \\
f_{\eta'}^{\q} & f_{\eta'}^{\s} & f_{\eta'}^{\c} \\
f_{\eta_{\c}}^{\q} & f_{\eta_{\c}}^{\s} & f_{\eta_{\c}}^{\c}
\end{array} \right)
 = U(\phi,\theta_y,\theta_{\c}) \, {\rm diag}(f_{\q},f_{\s},f_{\c})
\end{equation}
with the following parametrization of the transformation matrix
which now involve three angles
\begin{eqnarray}
U(\phi,\theta_y,\theta_{\c}) &:=& \left( \begin{array}{ccc}
\cos \phi & - \sin\phi & - \theta_{\c} \, \sin\theta_y \\
\sin \phi & \phantom{-{}}\cos\phi 
& \phantom{-{}} \theta_{\c} \, \cos\theta_y \\
- \theta_{\c} \, \sin(\phi-\theta_y) &
- \theta_{\c} \, \cos(\phi-\theta_y) & 1 
\end{array} \right) .
\end{eqnarray}
We neglect terms of order $\theta_{\c}^2$ 
since the mixing between $\eta$--$\eta'$ 
and $\eta_{\c}$ is an effect of the order of the inverse $\eta_{\c}$
mass, $M_{\eta_{\c}}$, squared; therefore 
we have $U U^\dagger = 1 + O(\theta_{\c}^2)$.
The two new mixing angles $\theta_{\c}$ and $\theta_y$ are related to
the ratios 
$f_{\eta'}^{{\c}}/f_{{\c}}$ and $f_{\eta}^{{\c}}/f_{\eta'}^{{\c}}$.
We have 
$f_{\eta}^{{\c}}= - f_{{\c}} \,\theta_{{\c}}\, \sin\theta_y$,
$f_{\eta'}^{{\c}} = f_{{\c}} \, \theta_{{\c}} \, \cos\theta_y$ and
$f_{\eta_{{\c}}}^{{\c}} \equiv f_{\eta_{{\c}}} = f_{{\c}} $, 
in accord with the definition utilized in
\cite{AlGr97}. $f_{\eta_{\c}}$ is the usual $\eta_{\c}$ decay
constant;
for its value we use the approximation $f_{\eta_{\c}} \simeq f_{J/\psi}$
and double the experimental error of $f_{J/\psi}$ for numerical
calculations ($f_{\eta_{\c}} = 405 \pm 30$~MeV, see \cite{FeKr97a}).
Accordingly, the mass matrix in the $\qbq$--$\sbs$--${\cbc}$ basis reads
 ($i,j=\q,\s,\c$),
\begin{equation}
{\mathcal M}_{\q\s{\c}}^2 =
 U^\dagger(\phi,\theta_y,\theta_{{\c}})  \, 
    {\rm diag}(M_{\eta\phantom{{}'}}^2,M_{\eta'}^2,M_{\eta_{{\c}}}^2) \,
     U(\phi,\theta_y,\theta_{{\c}}) 
              \ . 
\label{mij}
\end{equation}
On the other hand, generalizing Eq.~(\ref{qsmass}) and using
the abbreviations (\ref{miidef}), (\ref{ydef}) and (\ref{aadef})
introduced in Sect.~\ref{theorysect}, we may write
the mass matrix as follows
\begin{eqnarray}
{\mathcal M}_{\q\s\c}^2 =  
\left( \begin{array}{ccc} 
m_{\q\q}^2 & 0 & 0\\ 0 & m_{\s\s}^2 & 0 \\
0 & 0 & m_{\c\c}^2 
\end{array} \right)
+ 
\left( \begin{array}{ccc} 
2 a^2 & y \, \sqrt2 a^2 & z \, \sqrt2 a^2 \\
y \, \sqrt2 a^2 & y^2 \, a^2 & yz \, a^2 \\
z \, \sqrt2 a^2 & yz \, a^2 & z^2 \, a^2 
\end{array} \right) \ . 
\label{massansatz2}
\end{eqnarray}
On exploiting again the divergences of the axial
vector currents and the properties of the mass matrix a
number of consequences follows from which all new parameters appearing in
Eq.~(\ref{massansatz2}) can be fixed
\begin{equation}
    z = f_{\q}/f_{\c} \ , \qquad  \theta_y = \theta_8 \ , \qquad
        M^2_{\eta_{\c}} = m_{\c\c}^2 + z^2 \, a^2 
\label{par1}
\end{equation}
and
\begin{equation}
       \theta_{\c} = - z \, \sqrt{2+y^2} \,
        \frac{a^2}{M^2_{\eta_{\c}}}
\label{par2}
\end{equation}
with $a^2$ as given in Eq.~(\ref{aavalue}).
Using the phenomenological parameter values quoted in
Table~\ref{comptable1}, we
find the following numerical results:
$z=0.35\pm 0.03$; $\theta_{\c} = -1.0^\circ\pm 0.1^\circ$. Since 
$z^2a^2= 0.03 \gev^2$
we have $M_{\eta_{\c}}^2=m_{\c\c}^2$ to a very good approximation.
The charm decay constants of the $\eta$ and the $\eta'$ take the
values
\begin{equation}
f^{\c}_{\eta} = -(2.4\pm 0.2) \mev \qquad
f^{\c}_{\eta'} = -(6.3\pm 0.6) \mev  
\end{equation}
Their values are in rough agreement with the results presented in
\cite{Ali97b,Petrov,Chao}
 but in dramatic conflict with the values quoted in
\cite{Cheng,Halperin}. $f_{\eta'}^{\c}$ lies well within the bound
estimated in \cite{FeKr97b}. Our analysis supports
the conclusions drawn in \cite{Ali97b} that the charm content
of the
$\eta'$ is not the solution for the abnormally large $B\to
K\eta'$ decay width, the explanation of which remains an open problem.

Using the above values for the mixing angles $\phi$, $\theta_y$ 
and $\theta_{\c}$, we find for the quark content of the physical mesons
\begin{eqnarray}
|\eta\phantom{{}'} \rangle &=&   
       0.77\phantom{0} \, |\eta_{\q}\rangle  
     - 0.63\phantom{0} \, |\eta_{\s}\rangle 
     - 0.006\, |\eta_{\c 0}\rangle \nonumber\\
|\eta' \rangle &=& 
       0.63\phantom{0} \, |\eta_{\q}\rangle 
     + 0.77\phantom{0} \, |\eta_{\s}\rangle 
     - 0.016\, |\eta_{\c 0}\rangle \nonumber\\
|\eta_{\c} \rangle &=& 
       0.015 \, |\eta_{\q}\rangle 
     + 0.008 \, |\eta_{\s}\rangle 
     + \phantom{1.000} \, |\eta_{\c 0}\rangle .
\label{admixtures}
\end{eqnarray}
The charm admixtures to the $\eta$ and $\eta'$ are somewhat smaller
than estimated in \cite{Fritzsch} but slightly larger than
quoted in \cite{Chao}.
A possible test for the $\eta_{c0}$ content is provided by
the radiative $J/\psi$ decays. For the decays $J/\psi \to \eta\gamma,
\eta'\gamma$ we used already the action of the gluons as 
described by the matrix elements of the anomaly
(note that $\theta_8=\theta_y$). Since the $\eta'$ has
the $\eta_{c0}$ content $\theta_c \, \cos\theta_8$ while the
$\eta_{c0}$ content of $\eta_c$ is practically one, we expect
\begin{eqnarray}
\frac{\Gamma[J/\psi \to \eta'\gamma]}{\Gamma[J/\psi\to \eta_c\gamma]}
&=& \theta_c^2 \, \cos^2\theta_8 \,
\left(\frac{k_{\eta'}}{k_{\eta_c}}\right)^3
= \left(\frac{\langle 0 | \frac{\alpha_s}{4\pi} \, G \tilde G
    |\eta'\rangle}{\sqrt2 \, f_{\eta_{\c}} \, M_{\eta_{\c}}^2 } \right)^2 \,
\left(\frac{k_{\eta'}}{k_{\eta_c}}\right)^3
\label{etacetap}
\end{eqnarray}
The experimental number for this ratio, $0.33 \pm 0.10$ \cite{PDG96},
gives us another -- admittedly less reliable -- determination of
the charm admixture in $\eta'$. The result $|\theta_c\cos\theta_8|
=0.014 \pm 0.002$ is in good agreement with the number contained in
Eq.~(\ref{admixtures}).

\section{Summary}
\label{summsec}
In the description of $\eta$ - $\eta'$ mixing there are five parameters 
involved, the mixing angle of the particle states and four decay
constants. Motivated by the observation of nearly ideal mixing
in vector and tensor particles we take as our basis the
states according to their quark flavor compositions. Our central assumption 
is then that in this particular basis the mixing of the decay 
constants follows that of state mixing. This new mixing 
scheme is very restrictive. It fixes the structure of the mass
matrix and predicts the mixing angle and the four decay constants up
to first order in flavor symmetry breaking:

i) The four decay constants are immediately reduced to two constants 
$f_{\q} $ and $f_{\s}$ and a single angle $\phi$ that is identical to 
the state mixing angle and describes the deviation from ideal mixing.

ii) The divergences of the axial vector currents provide us 
with a mass matrix quadratic in the particle masses with 
off diagonal elements entirely determined by the anomaly. 
The old problem of quadratic versus linear mass matrices has 
found its answer.

iii) The ratio of matrix elements of the anomaly are equal 
to the inverse ratio of the corresponding decay constants 
corresponding to the 
states of our basis. Using this result the mixing angle $\phi$ 
can be calculated from $f_{\q}/f_{\s}$.

iv) $\sudrei$ relations fix $f_{\q}$ and $f_{\s}$ in terms of
$f_{\pi}$ and $f_K$ to first order in flavor symmetry breaking, and 
fix those parts of the mass matrix which contain the current 
quark masses in terms of $M_{\pi}^2$ and $M_K^2$ . The decay constants
obtained this way obey the requirements 
of chiral perturbation theory which are known up to order 
$1/N_c$ corrections.

v) With these ingredients and by using the known masses of the 
physical states the mass matrix is over-determined. Although 
sizeable corrections to the flavor symmetry results could 
have been expected, the resulting parameter-free determination 
of the mixing angle and the decay constants is in reasonable 
agreement with a previous phenomenological analysis with 
unconstrained parameters.

vi) We performed a new analysis and determined phenomenologically
$\phi$ and $f_{\q}$ and $f_{\s}$  from several independent experiments. 
All results were consistent with each other. Thus, the weighted 
average value for the mixing angle is rather precise: we obtained 
$\phi = 39.3^{\circ} \pm 1^{\circ}$ which gives a single-octet 
mixing angle of $\theta = -15.4^{\circ}$ . For the angle $\theta_8$ 
which is responsible for the $\eta$, $\eta'$ ratio in radiative $J/\psi$ 
decays we found a value of $-21.2^{\circ}$.
The values for $f_{\q}$ and for $f_{\s}$ differ from the theoretical 
predictions (to first order of flavor symmetry breaking) only mildly.

vii) It is straightforward to generalize the new mixing scheme 
to include the mixing with the $\eta_{\c}$ which is of particular recent 
interest. Here the  decay constant $f_{\c}$ enters which we 
take equal to $f_{J/\psi}$. With this ingredient the $\cbc$ 
admixture of $\eta$ and $\eta'$ could be determined in magnitude 
and sign. For the magnitude nearly the same number follows 
from the observed ratio of $J/\psi$ decays to $\eta'$ and
$\eta_{\c}$ without invoking $f_{\c}$. For the decay constant 
$f^{\c}_{\eta'}$ we find a value of $ -(6.3 \pm 0.6)$ MeV.


\newpage

\begin{table}[bthp]
\begin{center}
\begin{tabular}{c| cccccc }
source & $f_{\q}/f_\pi$ & $f_{\s}/f_\pi$ & $\phi$ & $\theta$ & $y$
& 
$a^2$ $[$GeV$^2]$ \\
\hline\hline
theory (Sect.\ 2) & 
 $\protect\phantom{\pm}1.00$ & $\protect\phantom{\pm}1.41$ & 
 $\protect\phantom{\pm}42.4^\circ$ & $ -12.3^\circ$ & 
$\protect\phantom{\pm}0.78$ & $\protect\phantom{\pm}0.281$ \\
\hline
phenomenology (Sect.~\ref{phensect})   &
 $\protect\phantom{\pm}1.07$ & $\protect\phantom{\pm}1.34$ & 
 $\protect\phantom{\pm}39.3^\circ$ & $-15.4^\circ$ & 
 $\protect\phantom{\pm}0.81$ & $\protect\phantom{\pm}0.265$ \\
 & $\pm 0.02$ & $\pm 0.06$ & $\pm \protect\phantom{3}1.0^\circ$ & 
 $\pm\protect\phantom{1}1.0^\circ$ & $\pm 0.03$ & $\pm0.010 $ 
\end{tabular}
\end{center}
\caption{Theoretical (to first order of flavor symmetry
breaking) and phenomenological values of
mixing parameters. The parameter $y$ is calculated using
Eqs.~(\ref{aavalue}) and (\ref{yvalue}).}
\label{comptable1}
\end{table}

\begin{table}[hbtp]
\begin{center}
\begin{tabular}{c |cccc}
source & $f_8/f_{\pi}$ & $f_1/f_{\pi}$ & $\theta_8$ & $\theta_1$ \\
\hline 
theory (Sect.\ 2)         & 1.28 & 1.15 & $-21.0^\circ$ & $-2.7^\circ$
\\
\protect\cite{Leutwyler97} 
                          & 1.28 & 1.25 & $-20.5^\circ$ & $- 4^\circ$
\\
\protect\cite{FeKr97b}   
                          & 1.28 & 1.20 & $-22.2^\circ$ & $-9.1^\circ$ 
\\
phenomenology   
                          & 1.26 & 1.17 & $-21.2^\circ$ & $-9.2^\circ$
\end{tabular}
\caption{Comparison of various theoretical and phenomenological results
 for the decay parameters defined in Eq.~(\ref{newmix}), see text.}
\label{comptable2}
\end{center}
\end{table}





\begin{thebibliography}{10}
\bibitem{Fritzsch}
H.\ Fritzsch and J.D.\ Jackson,
\newblock Phys.\ Lett.\ {\bf 66B} (1977) 365.

\bibitem{Isgur}
N.\ Isgur,
\newblock Phys.\ Rev.\ {\bf D13} (1976) 122.

\bibitem{Gilman:1987ax}
F.~J. Gilman and R.~Kauffman,
\newblock Phys. Rev. {\bf D36} (1987) 2761.

\bibitem{dmitrasinovic97}
V.~Dmitrasinovic,
\newblock Phys.\ Rev.\ {\bf D56} (1997) 247.

\bibitem{Witten}
E.\ Witten, 
\newblock Nucl.\ Phys.\ {\bf B149} (1979) 285;
G.\ Veneziano,
\newblock ibid. {\bf B159} (1979) 213.  


\bibitem{Leutwyler97}
H.~Leutwyler,
\newblock Nucl. Phys. Proc. Suppl. {\bf 64} (1998) 223;
\newblock R. Kaiser, diploma work, Bern University 1997.

\bibitem{FeKr97b}
T.~Feldmann and P.~Kroll,
to be published in Eur.\ Phys.\ J.\ C \newblock (1998), hep-ph/9711231.

\bibitem{BL80}
S.~J. Brodsky and G.~P. Lepage,
\newblock Phys. Rev. {\bf D22} (1980) 2157.

\bibitem{JuWe98}
D.~U.~Jungnickel and C.~Wetterich,
\newblock Eur.\ Phys.\ J.\ {\bf C1} (1998) 669.

\bibitem{Cheng}
H.Y.\ Cheng and B.\ Tseng,
\newblock Phys. Lett. {\bf B415} (1997) 263.

\bibitem{Halperin}
I.\ Halperin and A.\ Zhitnitsky,
\newblock Phys. Rev. {\bf D56} (1997) 7247.
 
\bibitem{AlGr97}
A.~Ali and C.~Greub,
\newblock  Phys.\ Rev.\ {\bf D57} (1998) 2996. 

\bibitem{Ali97b}
A.~Ali, J.~Chay, C.~Greub and P.~Ko,
Phys.\ Lett.\ {\bf B424} (1998) 161.

\bibitem{exp}
CLEO collaboration, B.~H.~Behrens {\em et~al.},
Phys.\ Rev.\ Lett.{\bf 80} (1998) 3710.

\bibitem{BaFrTy95}
P.~Ball, J.~M. Frere and M.~Tytgat,
\newblock Phys. Lett. {\bf B365} (1996) 367.

\bibitem{Bramon97}
A.~Bramon, R.~Escribano and M.~D. Scadron,
\newblock Phys. Lett. {\bf B403}  (1997) 339;
A.~Bramon, R.~Escribano and M.~D. Scadron, hep-ph/9711229.

 
\bibitem{PDG96}
Particle Data Group, R.~M. Barnett {\em et~al.},
\newblock Phys. Rev. {\bf D54} (1996) 1.


\bibitem{Dumbrajs:1983jd}
O.~Dumbrajs {\em et~al.},
\newblock Nucl. Phys. {\bf B216} (1983) 277.

\bibitem{BSW} M.\ Wirbel, B.\ Stech and M.\ Bauer,
\newblock Z. Phys. {\bf C29} (1985) 637.


\bibitem{CLEODs}
CLEO collaboration, G.~Brandenburg {\em et~al.},
\newblock Phys. Rev. Lett. {\bf 75} (1995) 3804.

\bibitem{ape}
Serpukhov-CERN collaboration, W.~D. Apel {\em et~al.},
\newblock Phys. Lett. {\bf 83B} (1979) 131.

\bibitem{sta}
N.~R. Stanton {\em et~al.},
\newblock Phys. Lett. {\bf 92B} (1980) 353.

\bibitem{ams}
Crystal Barrel collaboration, C.~Amsler {\em et~al.},
\newblock Phys. Lett. {\bf B294} (1992) 451.

\bibitem{Novikov}
V.A.\ Novikov {\em et~al.},
\newblock Nucl.\ Phys.\ {\bf B165} (1980) 55.

\bibitem{CLEO97}
CLEO collaboration, J.~Gronberg {\em et~al.},
\newblock Phys. Rev. {\bf D57} (1998) 33;
L3 collaboration, M.~Acciarri {\em et~al.},
Phys.\ Lett.\ {\bf B418} (1998) 399.

\bibitem{FeKr97a}
T.~Feldmann and P.~Kroll,
\newblock Phys.\ Lett.\ {\bf B413} (1997) 410;
M.~Neubert and B.~Stech, hep-ph/9705292.

\bibitem{Petrov}
A.~Petrov,
 hep-ph/9712497. 

\bibitem{Chao}
K.~Chao, \newblock Nucl.\ Phys.\ {\bf B317} (1989) 597,
\newblock Nucl.\ Phys.\ {\bf 335} (1990) 101;
F.~Yuan and K.~Chao, \newblock Phys.\ Rev.\ {\bf D56} (1997) 2495.

\end{thebibliography}
\end{document}